\documentclass[11pt,a4paper]{article}
\usepackage[hyperref]{acl2021}
\usepackage{times}
\usepackage{latexsym}
\usepackage{subfigure}
\usepackage{graphicx}
\usepackage{multirow}
\usepackage{amsmath}
\usepackage{amsfonts}
\usepackage{enumitem}
\usepackage[absolute,overlay]{textpos}

\usepackage{microtype}

\aclfinalcopy 
\def\aclpaperid{717} 

\title{Personalized Transformer for Explainable Recommendation}

\author{Lei Li$^1$ \space\space Yongfeng Zhang$^2$ \space\space Li Chen$^1$ \\
	$^1$Hong Kong Baptist University, Hong Kong, China \\
	$^2$Rutgers University, New Brunswick, USA \\
	$^1$\texttt{\{csleili,lichen\}@comp.hkbu.edu.hk} \\ $^2$\texttt{yongfeng.zhang@rutgers.edu} \\}

\date{}

\begin{document}
\maketitle
\begin{abstract}
Personalization of natural language generation plays a vital role in a large spectrum of tasks, such as explainable recommendation, review summarization and dialog systems. In these tasks, user and item IDs are important identifiers for personalization. Transformer, which is demonstrated with strong language modeling capability, however, is not personalized and fails to make use of the user and item IDs since the ID tokens are not even in the same semantic space as the words. To address this problem, we present a PErsonalized Transformer for Explainable Recommendation (PETER\footnote{\url{https://github.com/lileipisces/PETER}}), on which we design a simple and effective learning objective that utilizes the IDs to predict the words in the target explanation, so as to endow the IDs with linguistic meanings and to achieve personalized Transformer. Besides generating explanations, PETER can also make recommendations, which makes it a unified model for the whole recommendation-explanation pipeline. Extensive experiments show that our small unpretrained model outperforms fine-tuned BERT on the generation task, in terms of both effectiveness and efficiency, which highlights the importance and the nice utility of our design.
\end{abstract}

\begin{textblock*}{10cm}(2cm,1cm) 
\underline{Published as a conference paper at ACL-IJCNLP 2021}
\end{textblock*}

\begin{figure*}
	\centering
	\subfigure[Standard Transformer model, where the user and the item have no contribution to each generation step.]{\includegraphics[scale=0.5]{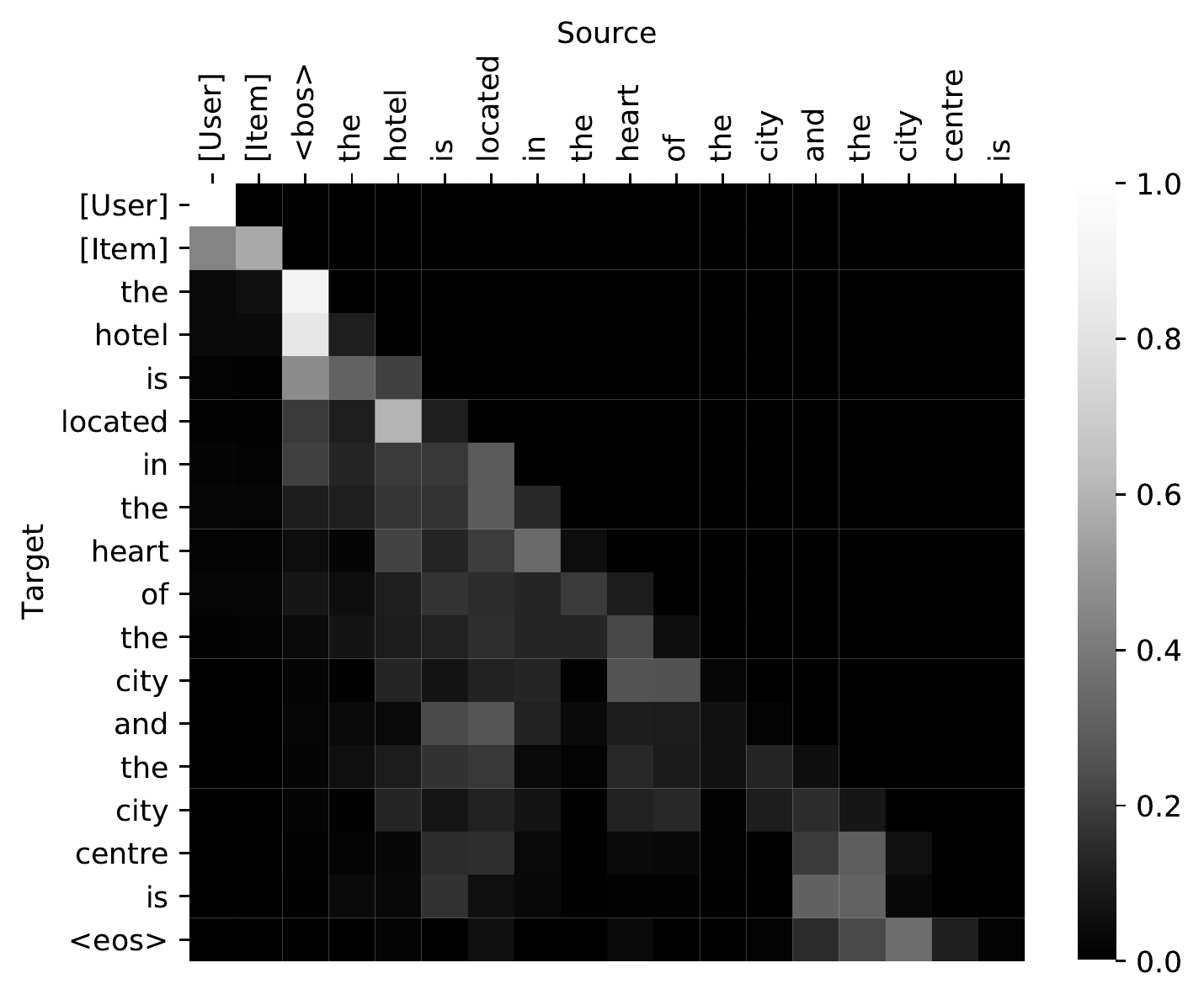}}
	\hspace{10mm}
	\subfigure[Our PETER model, where the user and item IDs play significant roles in the generation steps.]{\includegraphics[scale=0.5]{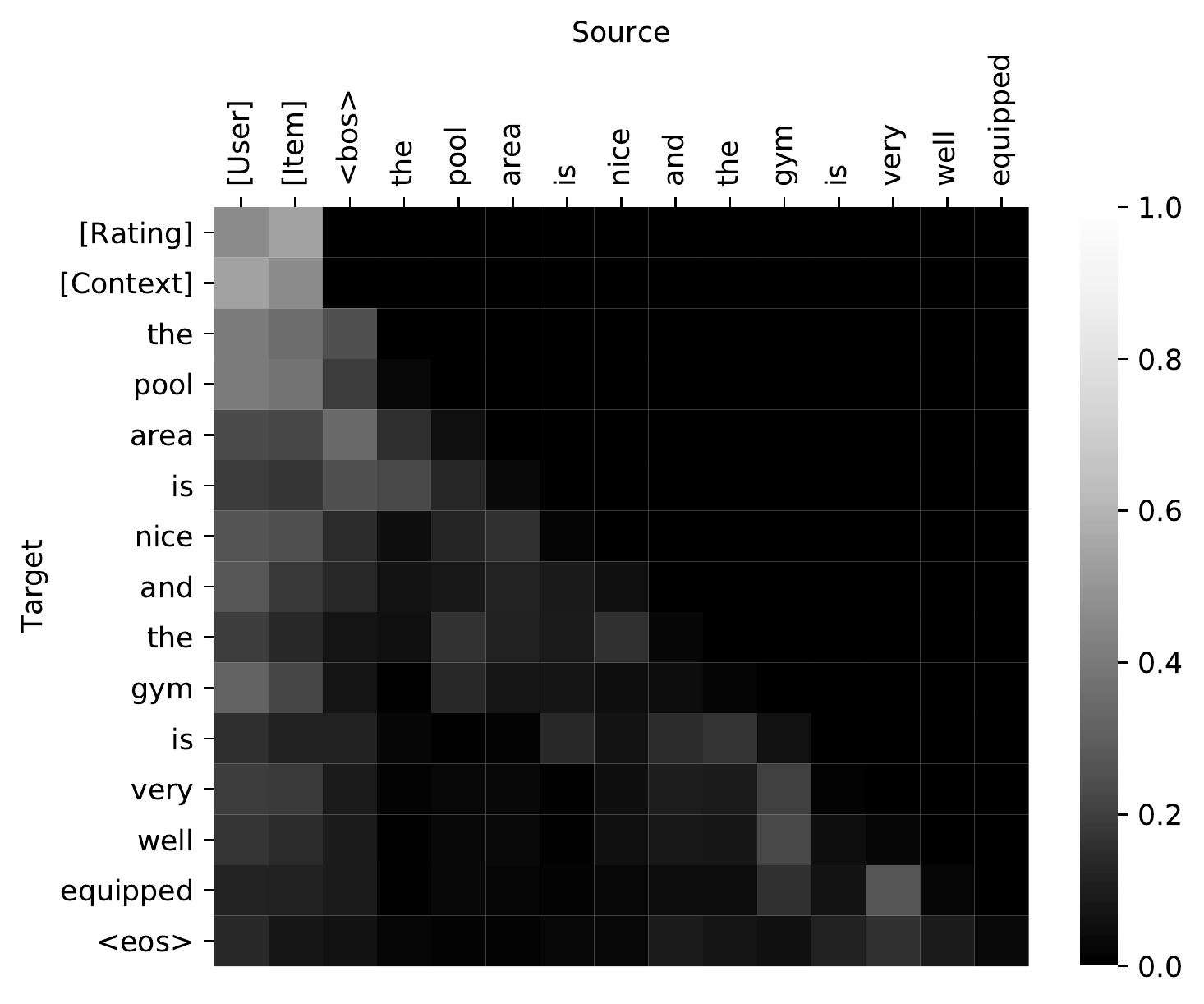}}
	\vspace{-5pt}
	\caption{Attention visualization of two models when generating an explanation for the same user-item pair (see the first two columns). They are both from the last attention layer, so the target sequences are offset by one position for better illustration. The larger the attention weights, the lighter the cells.}
	\label{fig:example}
	\vspace{-10pt}
\end{figure*}

\section{Introduction}

Recent years have witnessed the successful application of natural language generation.
Many of the applications in fact require certain degree of personalization, such as explainable recommendation \citep{SIGIR14-EFM, CIKM20-NETE, FTIR20-Survey}, review generation \citep{EACL17-Att2Seq}, review summarization \citep{AAAI19-USN}, and conversational systems \citep{CIKM18-SAUR, IJCAI20-ECR}.
In these tasks, user and item IDs that distinguish one user/item from the others are crucial to personalization.
For example, in recommender systems, different users may care about different item features (e.g., \textit{style} vs. \textit{quality}), and different items may have different characteristics (e.g., \textit{fashionable} vs. \textit{comfortable}).
The goal of explainable recommendation \cite{FTIR20-Survey} is to provide an explanation to a user for a recommended item, so as to justify how the recommendation might match his/her interests.
That is, given a pair of user ID and item ID, the system needs to generate an explanation, such as ``\textit{the style of the jacket is fashionable}'' (see the last column of Table \ref{table:case} for more examples).

Transformer \citep{NIPS17-Transformer}, whose strong language modeling ability has been demonstrated on a variety of tasks \citep{18-GPT, NAACL19-BERT, NeurIPS20-GPT3}, however, is relatively under-explored for \textit{personalized} natural language generation.
Since IDs and words are in very different semantic spaces, it would be problematic to directly put them together for attention learning, because by doing so, the IDs are treated as words, but the IDs appear far less frequently than the words.
For example, a paragraph of review (and thus hundreds of words) on e-commerce platform only corresponds to a single pair of user ID and item ID.
As such, the IDs may be regarded as out-of-vocabulary tokens, to which the model is insensitive.
As shown in Fig. \ref{fig:example}(a), when generating an explanation for a user-item pair, standard Transformer relies heavily on the special $<$\textit{bos}$>$ token instead of the user or the item.
This would result in identical explanations over different user-item pairs (see USR score in Table \ref{table:explanation}), deviating from our personalization goal.

To address this problem, we bridge IDs and words by designing an elegant task called \textit{context prediction}, which maps IDs onto words to be generated by the explanation task.
This in some way resembles one's drafting-polishing process, where by predicting some words the \textit{context prediction} task does the job of drafting.
Then, the \textit{explanation generation} task polishes these words so as to form a readable sentence.
Meanwhile, we demonstrate that conducting \textit{recommendation} task on the same model is also feasible, so we name it PETER, which stands for PErsonalized Transformer for Explainable Recommendation.
As we can see in Fig. \ref{fig:example}(b), when PETER generates an explanation for the same user-item pair, it can utilize the information of both the user and the item, which illustrates the effectiveness of our \textit{context prediction} task.

In addition, PETER is flexible to incorporate item features that can help to guide its generation.
This can be very useful when, for instance, a user proactively asks the system to explain certain feature(s) of a recommendation \citep{CIKM20-NETE}, e.g., \textit{price}.
Then, we would expect the model to generate a targeted explanation, such as ``\textit{great jacket, especially for the price}''.
PETER is a small unpretrained Transformer with only 2 layers, yet it outperforms a fine-tuned BERT \citep{EMNLP19-ACMLM} on most metrics by a large margin, and takes less time to train, as shown in our experiments.
This manifests the superiority of our model.

In summary, our key contributions are:
\begin{itemize}
	\item We propose PETER that makes recommendation and generates explanation simultaneously based on user and item IDs for explainable recommendation.
	To the best of our knowledge, we are the first to enable Transformer with personalized natural language generation.
	\item We evaluate the generated explanations on not only text quality metrics (such as BLEU and ROUGE), but also metrics that particularly focus on explainability from the angle of item features.
	Extensive experiments show that our model can outperform state-of-the-art baselines on large datasets.
	\item Our solution sheds light on a broader scope of fields that also need personalization (e.g., personalized conversational systems).
	In addition, it points out a way for Transformer to deal with heterogeneous inputs, e.g., text and images in multimodal artificial intelligence.
\end{itemize}

\section{Related Work}

\textbf{Explainable recommendation} \citep{SIGIR14-EFM, FTIR20-Survey} has been studied from two major perspectives: human-computer interaction and machine learning.
The former \citep{IJHCS14-HCI, IUI17-HCI, chen2019user} investigates how people perceive different styles of explanations, while the latter provides explanations by designing new explainable recommendation algorithms, to which our work is more related.
There exist various types of explanation styles, such as pre-defined templates \citep{SIGIR14-EFM, JIIS20-CAESAR}, ranked sentences \citep{AAAI19-DER, SIGIR21-EXTRA}, image visualizations \citep{SIGIR19-VECF}, knowledge graph paths \citep{MDPI18-KG, SIGIR19-PGPR, SIGIR20-KGAT, CIKM20-CAFE}, reasoning rules \citep{shi2020neural, chen2021neural, zhu2021faithfully}, etc., among which, recently, generated natural language explanations \citep{EMNLP19-ACMLM, CIKM20-NETE} have received much attention, mainly owing to the advancement of natural language generation technology and the availability of textual data on recommendation platforms such as e-commerce.
However, previous works mostly rely on recurrent neural networks (RNN), e.g., LSTM \citep{Neural97-LSTM} and GRU \citep{EMNLP14-GRU}, leaving the potentially more effective Transformer under-explored, which motivates this work.

\textbf{Transformer} \citep{NIPS17-Transformer} was first brought to machine translation with the architecture of encoder-decoder.
Later works \citep{ICLR18-Decoder, NAACL19-BERT} show that it remains effective, even when the encoder or the decoder is removed, reducing nearly half of the parameters.
Under the paradigm of pre-training plus fine-tuning, Transformer's effectiveness has been confirmed on a wide range of tasks, including both natural language understanding and generation \citep{18-GPT, NAACL19-BERT, NeurIPS19-UNILM}.
Particularly, it is able to perform novel tasks, e.g., arithmetic, after scaling up both the model and the training data \citep{19-GPT2, NeurIPS20-GPT3}.
However, it may not be friendly to researchers who do not possess large amounts of computing resources.
Instead, our work explores small unpretrained models, as they are computationally cheaper and more flexible when being adapted to new applications, e.g., personalized generation.

\textbf{Personalized generation} usually involves the IDs of users and items.
Previous approaches typically adopt multi-layer perceptron (MLP) to encode the IDs into a context vector, from which RNN can decode a word sequence.
This strategy can be found in many applications, such as review generation \citep{EACL17-Att2Seq}, tip generation \citep{SIGIR17-NRT} and explanation generation \citep{CIKM20-NETE}.
However, it does not fit Transformer that relies entirely on self-attention.
Probably because a proper solution to deal with heterogeneous inputs (i.e., IDs and words) is yet to be found, previous works with Transformer for personalized generation replace IDs with text segments, such as persona attributes \citep{AAAI20-Chatbot}, movie titles \citep{KDD20-Chatbot} and item features \citep{EMNLP19-ACMLM}, which are in the same semantic space as the word sequence to be generated.
In comparison, our solution is to design an effective task that can give the IDs linguistic meanings, thus connecting IDs with words.

\begin{figure}[t]
	\centering
	\includegraphics[scale=0.4]{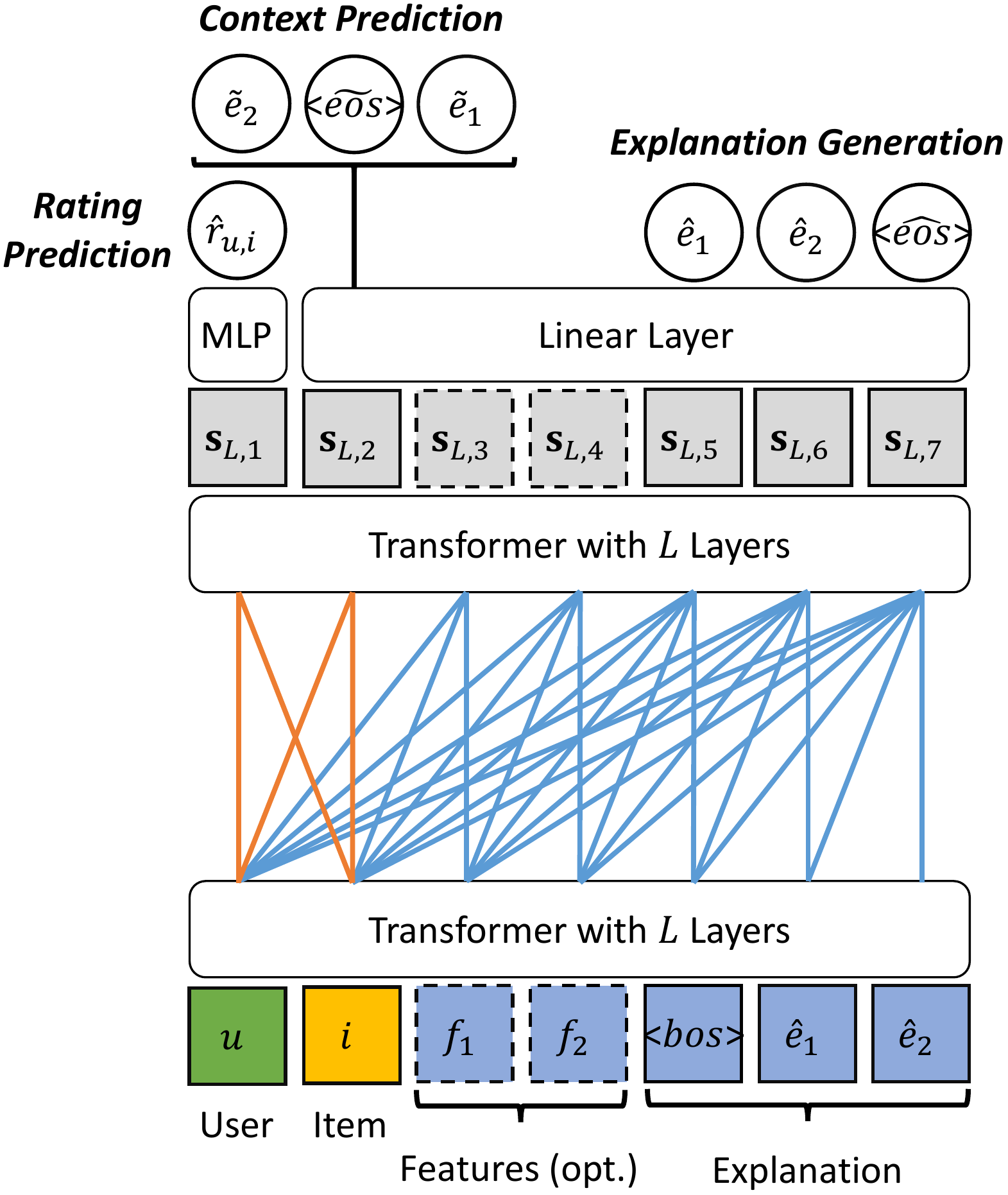}
	\caption{Our proposed model PETER that contains three tasks. The input features are optional.}
	\label{fig:model}
	\vspace{-10pt}
\end{figure}

\section{Problem Formulation}

The goal of our explanation task is to generate a natural language sentence $\hat{E}_{u, i}$ for a pair of user $u$ and item $i$ to justify why $i$ is recommended to $u$.
Meanwhile, our model PETER can also make recommendations by estimating a rating $\hat{r}_{u, i}$ that predicts $u$'s preference towards $i$.
At the testing stage, only user $u$ and item $i$ are used as inputs for producing both explanation and recommendation.
When item features $F_{u, i}$ are available, our model is flexible to incorporate them by simply concatenating them at the beginning of the explanation.
In this case, the features are also needed in the testing stage.
In the following, we will discuss both cases.

\section{Methodology}

In this section, we present the details of our model PETER.
First, we show how to encode different types of tokens in a sequence.
Then, we briefly review Transformer and introduce our revised attention masking matrix.
At last, we formulate the three tasks, i.e., explanation generation, context prediction and recommendation, and integrate them into a multi-task learning framework.

\subsection{Input Representation}

We first introduce our way to encode heterogeneous inputs into vector representations.
As shown in Fig. \ref{fig:model}, the input to our model is a sequence, consisting of user ID $u$, item ID $i$, features $F_{u, i}$, and explanation $E_{u, i}$.
The user and the item serve for the purpose of personalization, i.e., aiming to make the generated explanation reflect both the user's interests and the item's attributes.
The features can guide the model to talk about certain topics.
For instance, a conversational recommender system may explain a recommendation's specialty to the user with the goal of knowing more about his/her preference \citep{IJCAI20-ECR}.
Since the features are not always available, in our experiments we test both cases (with and without them).
When they are available, the input sequence can be represented as $S = [u, i, f_1, \cdots, f_{\left| F_{u, i} \right|}, e_1, \cdots, e_{\left| E_{u, i} \right|}]$, where $f_1, \cdots, f_{\left| F_{u, i} \right|}$ are the features and $e_1, \cdots, e_{\left| E_{u, i} \right|}$ are the explanation's word sequence. $\left| F_{u, i} \right|$ denotes the number of features and $\left| E_{u, i} \right|$ is the number of words in the explanation.

\begin{figure}[t]
	\centering
	\includegraphics[scale=0.4]{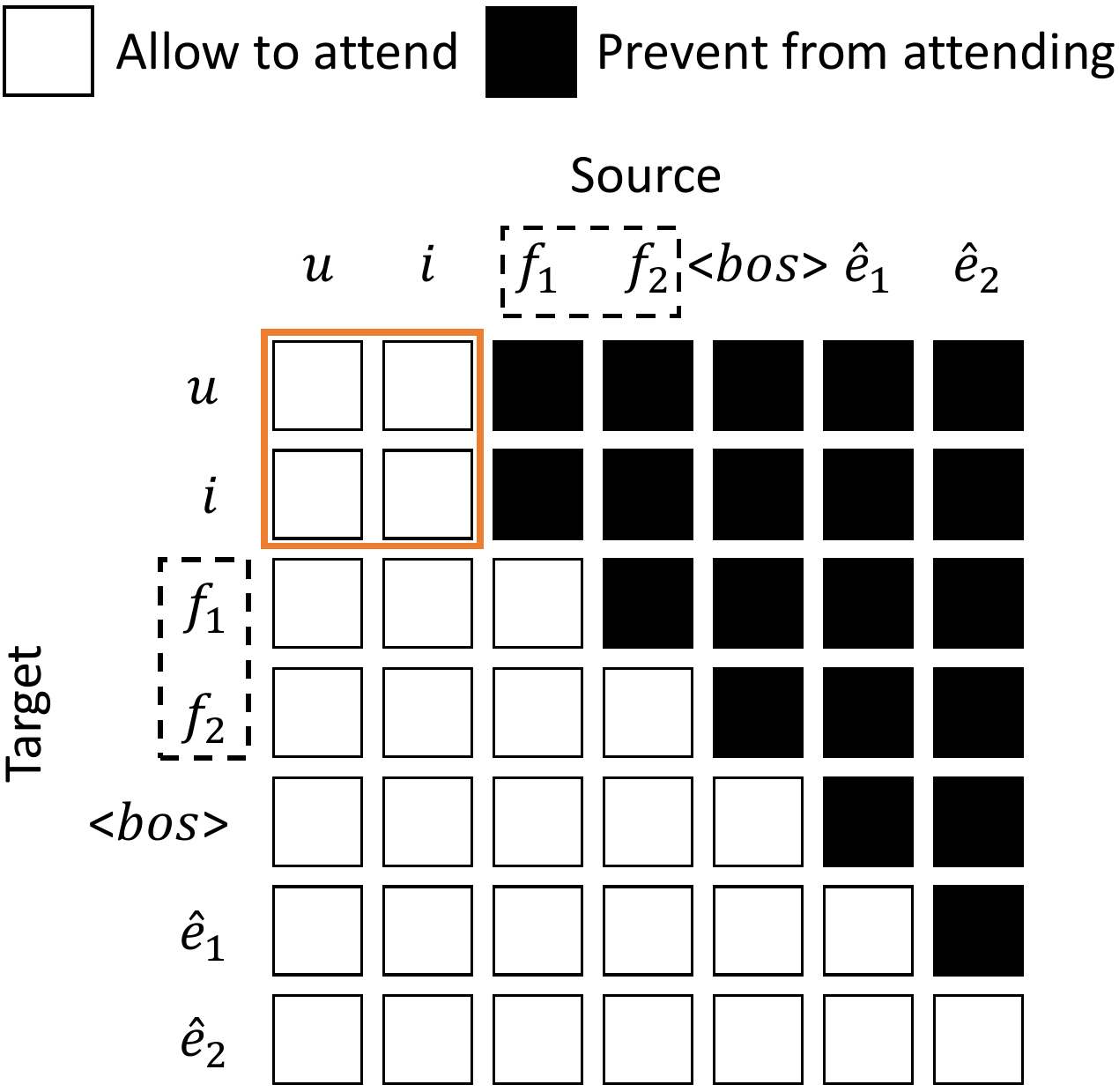}
	\caption{The attention masking used in our model that we call PETER masking. The orange box highlights its difference from the Left-to-Right masking.}
	\label{fig:mask}
	\vspace{-15pt}
\end{figure}

Clearly there are three types of tokens in the sequence $S$, i.e., users, items, and words (including features), for which we prepare three sets of randomly initialized token embeddings $\mathbf{U}$, $\mathbf{I}$ and $\mathbf{V}$ respectively, besides the positional embeddings $\mathbf{P}$ that encode the position of each token in the sequence.
Notice that, we do not add users and items to the vocabulary $\mathcal{V}$, given that it costs more time to predict a word out of the huge amount of IDs (for example, millions of users and items in e-commerce).
After performing embedding look-up, we can obtain the sequence's token representation $[\mathbf{u}, \mathbf{i}, \mathbf{f}_1, \cdots, \mathbf{f}_{\left| F_{u, i} \right|}, \mathbf{e}_1, \cdots, \mathbf{e}_{\left| E_{u, i} \right|}]$ and its positional representation $[\mathbf{p}_1, \cdots, \mathbf{p}_{\left| S \right|}]$, where $\left| S \right|$ is the length of the sequence.
The input representation of the sequence is the addition of the corresponding token representation and positional representation, denoted as $\mathbf{S}_0 = [\mathbf{s}_{0, 1}, \cdots, \mathbf{s}_{0, \left| S \right|}]$.

\subsection{Transformer and Attention Masking}

To enable the three tasks, we show how to modify the attention masking mechanism in Transformer \citep{NIPS17-Transformer}.
Transformer consists of $L$ identical layers, each of which is composed of two sub-layers: multi-head self-attention and position-wise feed-forward network.
The $l$-th layer encodes the previous layer's output $\mathbf{S}_{l - 1}$ into $\mathbf{S}_l$, where $l \in [1, L]$.
In the multi-head self-attention sub-layer, the computation of each attention head is also identical, and among the $H$ heads of the $l$-th layer, the $h$-th head $\mathbf{A}_{l, h}$ is computed as follows:
\begin{equation}
	\begin{aligned}
		\mathbf{A}_{l, h} & = \text{softmax} (\frac{\mathbf{Q}_{l, h} \mathbf{K}_{l, h}^\top}{\sqrt{d}} + \mathbf{M}) \mathbf{V}_{l, h} \\
		\mathbf{Q}_{l, h} & = \mathbf{S}_{l - 1} \mathbf{W}_{l, h}^Q, \mathbf{K}_{l, h} = \mathbf{S}_{l - 1} \mathbf{W}_{l, h}^K, \\
		\mathbf{V}_{l, h} & = \mathbf{S}_{l - 1} \mathbf{W}_{l, h}^V \\
		\mathbf{M} & =
		\begin{cases}
			0, & \text{Allow to attend} \\
			- \infty, & \text{Prevent from attending}
		\end{cases}
	\end{aligned}
\end{equation}
where $\mathbf{S}_{l - 1} \in \mathbb{R}^{\left| S \right| \times d}$ is the $(l - 1)$-th layer's output, $\mathbf{W}_{l, h}^Q, \mathbf{W}_{l, h}^K, \mathbf{W}_{l, h}^V \in \mathbb{R}^{d \times \frac{d}{H}}$ are projection matrices, $d$ denotes the dimension of embeddings, and $\mathbf{M} \in \mathbb{R}^{\left| S \right| \times \left| S \right|}$ is the attention masking matrix.

Each element in $\mathbf{M}$ controls whether a token in the sequence can attend to another.
For example, in the unidirectional left-to-right language model \citep{18-GPT}, the lower triangular part of $\mathbf{M}$ is set to $0$ and the remaining part $- \infty$, so as to allow each token to attend to past tokens (including itself), but prevent it from attending to future tokens.
We call it \textit{Left-to-Right Masking}.
As our model is not limited to the left-to-right explanation generation task, we modify the masking mechanism to accommodate the other two tasks (i.e., context prediction and recommendation).
As shown in Fig. \ref{fig:mask}, the first two tokens $u$ and $i$ in the sequence can attend to each other, because both context prediction and recommendation tasks need them.
To echo our model, we name it \textit{PETER Masking}.

\subsection{Explanation and Recommendation} \label{sec:task}

In the following, we perform the three tasks, after obtaining the sequence's final representation $\mathbf{S}_L = [\mathbf{s}_{L, 1}, \cdots, \mathbf{s}_{L, \left| S \right|}]$ from Transformer.
The key challenge lies in the personalization of explanation generation task, for which we design the context prediction task.
For both tasks, we apply a linear layer to the final representation of each token to map it into a $\left| \mathcal{V} \right|$-sized vector.
As an example, after passing through this layer, $\mathbf{s}_{L, t}$ becomes $\mathbf{c}_t$:
\begin{equation}
	\mathbf{c}_t = \text{softmax} (\mathbf{W}^v \mathbf{s}_{L, t} + \mathbf{b}^v)
\end{equation}
where $\mathbf{W}^v \in \mathbb{R}^{\left| \mathcal{V} \right| \times d}$ and $\mathbf{b}^v \in \mathbb{R}^{\left| \mathcal{V} \right|}$ are weight parameters.
The vector $\mathbf{c}_t$ represents the probability distribution over the vocabulary $\mathcal{V}$, from which a word $e$ with probability $c_t^e$ can be sampled.

\paragraph{Explanation Generation:}
We adopt the Negative Log-Likelihood (NLL) as the explanation task's loss function, and compute the mean of user-item pairs in the training set:
\begin{equation}
	\mathcal{L}_e = \frac{1} {\left| \mathcal{T} \right|} \sum_{(u, i) \in \mathcal{T}} \frac{1}{\left| E_{u, i} \right|} \sum_{t = 1}^{\left| E_{u, i} \right|} - \log c_{2 + \left| F_{u, i} \right| + t}^{e_t}
	\label{eq:explanation}
\end{equation}
where $\mathcal{T}$ denotes the training set.
The probability $c_t^{e_t}$ is offset by $2 + \left| F_{u, i} \right|$ positions because the explanation is placed at the end of the sequence, and $\left| F_{u, i} \right| = 0$ when the features are unavailable.

At the testing stage, along with $u$, $i$, and $F_{u, i}$ (if available), we feed the model a special begin-of-sequence token $<$\textit{bos}$>$.
From its resulting probability distribution $\mathbf{c}_{<bos>}$, the model can predict a word.
For simplicity, among the many decoding methods, we opt for greedy decoding that samples the word with the largest probability.
Then we can concatenate this predicted word at the end of the sequence to form a new input sequence for generating another word.
We do this repeatedly until the model produces a special end-of-sequence token $<$\textit{eos}$>$, or the generated explanation $\hat{E}_{u, i}$ reaches a pre-defined length.

\paragraph{Context Prediction:}
As discussed earlier, when there is only one task of explanation generation, Transformer fails to make use of user ID and item ID, resulting in identical sentences.
To address this issue, we design this task to map the IDs onto the words in the explanation, so as to build a connection between them.
Since the first two positions ($u$ and $i$) of the sequence are allowed to attend to each other, both of their final representations absorb the information of the user and the item.
Thus, we can use either of them to perform this task.
Here, we use the 2nd one for better illustration in Fig. \ref{fig:model}.
Again, we adopt NLL as the loss function:
\begin{equation}
	\mathcal{L}_c = \frac{1} {\left| \mathcal{T} \right|} \sum_{(u, i) \in \mathcal{T}} \frac{1}{\left| E_{u, i} \right|} \sum_{t = 1}^{\left| E_{u, i} \right|} - \log c_2^{e_t}
\end{equation}
where the difference from Eq. \eqref{eq:explanation} is that all predicted words are from the 2nd position, which is why they are not sequentially ordered (see Fig. \ref{fig:model}).

\paragraph{Rating Prediction:}
Recommendation can be seen as a prediction problem \citep{chen2021neural} where the goal is to predict a score $\hat{r}_{u, i}$ based on the IDs of user $u$ and item $i$.
As both $u$ and $i$ in the sequence can attend to each other, their final representations capture the interaction between them.
Next, we map the 1st representation $\mathbf{s}_{L, 1}$ into a scalar (because the 2nd one is used for context prediction).
To this end, we employ multi-layer perceptron (MLP) with one hidden layer as follows:
\begin{equation}
	\hat{r}_{u, i} = \mathbf{w}^r \sigma(\mathbf{W}^r \mathbf{s}_{L, 1} + \mathbf{b}^r) + b^r
\end{equation}
where $\mathbf{W}^r \in \mathbb{R}^{d \times d}$, $\mathbf{b}^r \in \mathbb{R}^d$, $\mathbf{w}^r \in \mathbb{R}^{1 \times d}$ and $b^r \in \mathbb{R}$ are weight parameters, and $\sigma(\cdot)$ is the sigmoid function.
Therefore, it can be seen that it is feasible to do both recommendation and explanation on Transformer.
As recommendation is not the key focus of this paper, we leave its improvement in the future work.
For this task, we use Mean Square Error (MSE) as the loss function:
\begin{equation}
	\mathcal{L}_r = \frac{1} {\left| \mathcal{T} \right|} \sum_{(u, i) \in \mathcal{T}} (r_{u, i} - \hat{r}_{u, i})^2
\end{equation}
where $r_{u, i}$ is the ground-truth rating.

\paragraph{Multi-task Learning:}
At last, we integrate the three tasks into a multi-task learning framework whose objective function is defined as:
\begin{equation}
	\mathcal{J} = \min_{\Theta} (\lambda_e \mathcal{L}_e + \lambda_c \mathcal{L}_c + \lambda_r \mathcal{L}_r)
\end{equation}
where $\Theta$ denotes all the trainable parameters in the model, and $\lambda_e$, $\lambda_c$ and $\lambda_r$ are regularization weights that balance the learning of different tasks.
In this way, the model can be trained efficiently in an end-to-end manner.

\section{Experimental Setup}

\subsection{Datasets}

For experimentation, we adopt three publicly available explainable recommendation datasets, and their data splits \citep{CIKM20-NETE}.
During the splitting process, each dataset is randomly divided into training, validation and testing sets with ratio 8:1:1 for 5 times, and the training set holds at least one record for each user and each item.
The three datasets are respectively from TripAdvisor (hotel), Amazon (movies \& TV) and Yelp (restaurant).
Each record in the datasets is comprised of a user ID, an item ID, a rating, an explanation, and a feature.
The explanations are sentences extracted from user reviews.
Each explanation contains at least one item feature, e.g., \textit{bedroom}, which ensures the explanation quality.
Statistics of the datasets are shown in Table \ref{table:dataset}.
We can see that Yelp is much larger than the other two in terms of size, making it closer to the real-world situation where there are millions of users and items.

\begin{table}
	\centering
	\small
	\begin{tabular}{l|r|r|r}
		\hline
		& Yelp & Amazon & TripAdvisor \\
		\hline
		\#users & 27,147 & 7,506 & 9,765 \\
		\#items & 20,266 & 7,360 & 6,280 \\
		\#records & 1,293,247 & 441,783 & 320,023 \\
		\#features & 7,340 & 5,399 & 5,069 \\
		\#records / user & 47.64 &58.86&32.77 \\
		\#records / item & 63.81 &60.02&50.96 \\
		\#words / exp & 12.32 &14.14&13.01 \\
		\hline
		\multicolumn{4}{l}{* \textbf{exp} denotes \textbf{explanation}.}
	\end{tabular}
	\caption{Statistics of the three datasets.}
	\label{table:dataset}
\end{table}

\subsection{Evaluation Metrics}

To evaluate the recommendation performance, we adopt two commonly used metrics: Root Mean Square Error (\textbf{RMSE}) and Mean Absolute Error (\textbf{MAE}).
As to explanation performance, we measure the generated explanations from two main perspectives: text quality and explainability.
For the former, we adopt \textbf{BLEU} \citep{ACL02-BLEU} in machine translation and \textbf{ROUGE} \citep{TS04-ROUGE} in text summarization, and report BLEU-1 and BLEU-4, and Precision, Recall and F1 of ROUGE-1 and ROUGE-2.
Though being widely used, BLUE and ROUGE are not flawless.
For example, it is difficult for them to detect the problem of identical sentences generated by Transformer.
These identical sentences might not be used as explanations, because they are less likely to well explain the special property of different recommendations.
To quantitatively measure how severe the problem is, we adopt \textbf{USR} that computes the Unique Sentence Ratio of generated sentences \citep{CIKM20-NETE}.

Text quality, however, is not equal to explainbility.
In the case of explainable recommendation, users may value more an explanation that justifies a recommendation's advantages on certain features \citep{CIKM20-NETE, EARS19-HSS}.
To this end, we adopt the other three metrics proposed by \citep{CIKM20-NETE}: Feature Matching Ratio (\textbf{FMR}), Feature Coverage Ratio (\textbf{FCR}) and Feature Diversity (\textbf{DIV}).
FMR measures whether a generated explanation contains the feature in the ground-truth.
FCR is computed as the number of distinct features contained in all the generated explanations, divided by the total number of features in the whole dataset.
DIV measures the intersection of features between any two generated explanations.

For RMSE, MAE and DIV, the lower, the better, while it is opposite for the rest of metrics.

\begin{table*}[t]
	\centering
	\small
	\setlength{\tabcolsep}{3.5pt}
	\begin{tabular}{r|lll|lllllllll}
		\hline
		\multirow{2}{*}{} & \multicolumn{3}{c|}{Explainability} & \multicolumn{9}{c}{Text Quality} \\ \cline{2-13}
		& FMR$\uparrow$ & FCR$\uparrow$ & DIV$\downarrow$ & USR$\uparrow$ & B1$\uparrow$ & B4$\uparrow$ & R1-P$\uparrow$ & R1-R$\uparrow$ & R1-F$\uparrow$ & R2-P$\uparrow$ & R2-R$\uparrow$ & R2-F$\uparrow$ \\ \hline \hline
		
		& \multicolumn{12}{|c}{Yelp} \\ \hline
		Transformer & 0.06 & 0.06 & 2.46 & 0.01 & 7.39 & 0.42 & \textbf{19.18} & 10.29 & 12.56 & 1.71 & 0.92 & 1.09 \\
		NRT & \underline{0.07} & 0.11 & \underline{2.37} & \underline{0.12} & \textbf{11.66} & \underline{0.65} & 17.69 & \underline{12.11} & \underline{13.55} & 1.76 & \underline{1.22} & \underline{1.33} \\
		Att2Seq & \underline{0.07} & \underline{0.12} & 2.41 & \textbf{0.13} & 10.29 & 0.58 & \underline{18.73} & 11.28 & 13.29 & \underline{1.85} & 1.14 & 1.31 \\
		PETER & \textbf{0.08}** & \textbf{0.19}** & \textbf{1.54}** & \textbf{0.13} & \underline{10.77} & \textbf{0.73}** & 18.54 & \textbf{12.20} & \textbf{13.77}** & \textbf{2.02}** & \textbf{1.38}** & \textbf{1.49}** \\ \hline
		ACMLM & 0.05 & \underline{0.31} & \textbf{0.95} & \textbf{0.95} & 7.01 & 0.24 & 7.89 & 7.54 & 6.82 & 0.44 & 0.48 & 0.39 \\
		NETE & \underline{0.80} & 0.27 & 1.48 & \underline{0.52} & \underline{19.31} & \underline{2.69} & \underline{33.98} & \underline{22.51} & \underline{25.56} & \underline{8.93} & \underline{5.54} & \underline{6.33} \\
		PETER+ & \textbf{0.86}** & \textbf{0.38}** & \underline{1.08} & 0.34 & \textbf{20.80}** & \textbf{3.43}** & \textbf{35.44}** & \textbf{26.12}** & \textbf{27.95}** & \textbf{10.65}** & \textbf{7.44}** & \textbf{7.94}** \\ \hline \hline
		
		& \multicolumn{12}{|c}{Amazon} \\ \hline
		Transformer & \underline{0.10} & 0.01 & 3.26 & 0.00 & 9.71 & 0.59 & 19.68 & 11.94 & 14.11 & 2.10 & 1.39 & 1.55 \\
		NRT & \textbf{0.12} & 0.07 & 2.93 & 0.17 & \textbf{12.93} & \underline{0.96} & \textbf{21.03} & \underline{13.57} & \textbf{15.56} & \underline{2.71} & \underline{1.84} & \underline{2.05} \\
		Att2Seq & \textbf{0.12} & \underline{0.20} & \underline{2.74} & \textbf{0.33} & 12.56 & 0.95 & \underline{20.79} & 13.31 & \underline{15.35} & 2.62 & 1.78 & 1.99 \\
		PETER & \textbf{0.12}** & \textbf{0.21} & \textbf{1.75}** & \underline{0.29} & \underline{12.77} & \textbf{1.17}** & 19.81 & \textbf{13.80} & 15.23 & \textbf{2.80} & \textbf{2.08}** & \textbf{2.20}** \\ \hline
		ACMLM & 0.10 & \textbf{0.31} & 2.07 & \textbf{0.96} & 9.52 & 0.22 & 11.65 & 10.39 & 9.69 & 0.71 & 0.81 & 0.64 \\
		NETE & \underline{0.71} & \underline{0.19} & \underline{1.93} & \underline{0.57} & \underline{18.76} & \underline{2.46} & \underline{33.87} & \underline{21.43} & \underline{24.81} & \underline{7.58} & \underline{4.77} & \underline{5.46} \\
		PETER+ & \textbf{0.77}** & \textbf{0.31}** & \textbf{1.20}** & 0.46 & \textbf{19.75}** & \textbf{3.06}** & \textbf{34.71}** & \textbf{23.99}** & \textbf{26.35}** & \textbf{9.04}** & \textbf{6.23}** & \textbf{6.71}** \\ \hline \hline
		
		& \multicolumn{12}{|c}{TripAdvisor} \\ \hline
		Transformer & 0.04 & 0.00 & 10.00 & 0.00 & 12.79 & 0.71 & 16.52 & \textbf{16.38} & 15.88 & 2.22 & \textbf{2.63} & \textbf{2.34} \\
		NRT & \underline{0.06} & 0.09 & \underline{4.27} & \underline{0.08} & 15.05 & 0.99 & 18.22 & 14.39 & 15.40 & 2.29 & 1.98 & 2.01 \\		
		Att2Seq & \underline{0.06} & \textbf{0.15} & 4.32 & \textbf{0.17} & \underline{15.27} & \underline{1.03} & \underline{18.97} & 14.72 & \underline{15.92} & \textbf{2.40} & 2.03 & \underline{2.09} \\		
		PETER & \textbf{0.07}** & \underline{0.13} & \textbf{2.95}** & \underline{0.08} & \textbf{15.96}** & \textbf{1.11}* & \textbf{19.07} & \underline{16.09} & \textbf{16.48}** & \underline{2.33} & \underline{2.17} & \underline{2.09} \\ \hline
		ACMLM & 0.07 & \textbf{0.41} & \textbf{0.78} & \textbf{0.94} & 3.45 & 0.02 & 4.86 & 3.82 & 3.72 & 0.18 & 0.20 & 0.16 \\
		NETE & \underline{0.78} & 0.27 & 2.22 & \underline{0.57} & \underline{22.39} & \underline{3.66} & \underline{35.68} & \underline{24.86} & \underline{27.71} & \underline{10.20} & \underline{6.98} & \underline{7.66} \\
		PETER+ & \textbf{0.89}** & \underline{0.35} & \underline{1.61} & 0.25 & \textbf{24.32}** & \textbf{4.55}** & \textbf{37.48}** & \textbf{29.21}** & \textbf{30.49}** & \textbf{11.92}** & \textbf{8.98}** & \textbf{9.24}** \\ \hline \hline
	\end{tabular}
	\caption{Performance comparison of the generation methods in terms of Explainability and Text Quality on three datasets. The methods are divided into two groups according to whether features are used or not. B1 and B4 stand for BLEU-1 and BLEU-4. R1-P, R1-R, R1-F, R2-P, R2-R and R2-F denote Precision, Recall and F1 of ROUGE-1 and ROUGE-2. BLEU and ROUGE are percentage values (\% symbol omitted for table clarity), while the others are absolute values. The best performing values are boldfaced, and the second best underlined. ** and * indicate the statistical significance over the second best baseline respectively for $p < 0.01$ and $p < 0.05$ via Student's t-test.}
	\label{table:explanation}
\end{table*}

\subsection{Compared Methods} \label{sec:methods}

We introduce baselines, first for explanation and then for recommendation. For the former, we divide the baselines into two groups, depending on whether the feature is used or not.

The following models leverage only user and item IDs to generate explanations (without feature). We denote our model without feature as PETER.

\begin{itemize}
	\item \textbf{Transformer} \citep{NIPS17-Transformer} performs the explanation generation task by treating user and item IDs as words.
	We also tested encoder-decoder Transformer, where the encoder encodes the IDs for the decoder to decode, but its results turned out to be the same, so we do not report it.
	\item \textbf{NRT} \citep{SIGIR17-NRT} can predict a rating and generate a tip simultaneously based on user and item IDs.
	We take the explanations in the datasets as tips.
	Moreover, we found that the model's problem of generating identical sentences (as reported in \citealp{CIKM20-NETE}) is caused by the L2 regularization in its original design.
	For fair comparison, we removed it.
	\item \textbf{Att2Seq} \citep{EACL17-Att2Seq} is a review generation approach and we take the explanations as reviews.
	This model has an attention module, but we found that it makes the generated content unreadable in the task.
	To be fair, we removed it as well.
\end{itemize}

When features are used, we denote our model as PETER+, and compare it with two recent models:

\begin{itemize}
	\item \textbf{ACMLM} \citep{EMNLP19-ACMLM} is a fine-tuned BERT \citep{NAACL19-BERT}, where an attention layer is introduced to encode the features from both the user and the item.
	By predicting masked tokens, this model can produce diverse sentences.
	\item \textbf{NETE} \citep{CIKM20-NETE}	is a tailored GRU \citep{EMNLP14-GRU} that incorporates a given feature into the decoding process to generate template-like explanations.
	It can also make recommendations.
\end{itemize}

For recommendation, besides NRT and NETE, we include another two traditional methods:

\begin{itemize}
	\item \textbf{PMF} \citep{NIPS08-PMF} is a standard probabilistic matrix factorization method that characterizes users and items by latent factors.
	\item \textbf{SVD++} \citep{KDD08-SVDpp} leverages a user's interacted items to enhance the latent factors.
\end{itemize}

\subsection{Implementation Details}

We train each model on the training set, tune the hyper-parameters on the validation set, and report the performance on the testing set.
The results are averaged on the 5 data splits.
We adopt the codes of ACMLM and NETE, and implement all the other methods.
For NRT, Att2Seq, NETE and our PETER and PETER+, we set the size of vocabulary to 20,000 by keeping the most frequent words.
We do not apply this to Transformer, otherwise users and items (regarded as words) may be filtered out.
We set both the number of context words and the length of explanations to 15, because the mean length of explanations is approximately 13 (see Table \ref{table:dataset}).
ACMLM adopts sub-words, so we do not apply the above two steps to it.
We reuse the other default settings of the baselines.

\begin{table}
	\centering
	\small
	\begin{tabular}{r|ccc}
		\hline
		\multicolumn{1}{c|}{} & Time & Epochs & Time/Epoch \\ \hline
		ACMLM & 97.0 & \textbf{3} & 32.3 \\
		PETER+ & \textbf{57.7} & 25 & \textbf{2.3} \\ \hline
	\end{tabular}
	\caption{Efficiency comparison of two Transformer-based models in terms of training minutes on the TripAdvisor dataset, tested on NVIDIA Tesla P40.}
	\label{table:efficiency}
\end{table}

\begin{table*}
	\centering
	\tiny
	\begin{tabular}{r|l|l}
		\hline
		& Top-15 Context Words & Explanation \\ \hline
		
		Ground-truth & & the \textbf{rooms} are spacious and the bathroom has a large tub \\
		PETER & $<$eos$>$ the and a \underline{pool} was with nice is very were to good in of & the \underline{pool} area is nice and the \underline{gym} is very well equipped $<$eos$>$ \\
		PETER+ & $<$eos$>$ the and a was \underline{pool} with to nice good very were is of in & the \underline{rooms} were clean and comfortable $<$eos$>$ \\ \hline
		
		Ground-truth & & beautiful \textbf{lobby} and nice bar \\
		PETER & $<$eos$>$ the and a was were separate \underline{bathroom} with \underline{shower} large very had in is & the \underline{bathroom} was large and the \underline{shower} was great $<$eos$>$ \\
		PETER+ & $<$eos$>$ the and a was \underline{bathroom} \underline{shower} with large in separate were \underline{room} very is & the \underline{lobby} was very nice and the \underline{rooms} were very comfortable $<$eos$>$ \\ \hline
	\end{tabular}
	\caption{Context words and explanations on two different cases as generated by our PETER and PETER+ on TripAdvisor dataset. The boldfaced words in the ground-truth are the key features. Generated features are underlined.}
	\vspace{-10pt}
	\label{table:case}
\end{table*}

For Transformer, PETER and PETER+, we set the embedding size $d$ to 512 and the dimension of feed-forward network to 2,048, following \citep{NIPS17-Transformer}, but the number of layers $L$ and attention heads $H$ are both 2.
For our models PETER and PETER+, we set the regularization weights $\lambda_e$, $\lambda_c$ and $\lambda_r$ to 1.0, 1.0 and 0.1, respectively.
We optimize the model via stochastic gradient descent \citep{TAMS51-SGD}, and apply gradient clipping \citep{ICML13-Clipping} with a threshold of 1.0.
The batch size is set to 128, and the learning rate 1.0.
At each epoch, we save the model if it achieves the lowest loss on the validation set, but when there is no improvement, we decrease the learning rate by a factor of 0.25.
When the latter happens for 5 times, we stop training and load the saved model for prediction.

\section{Results and Analysis}

\subsection{Quantitative Analysis on Explanations}

In Table \ref{table:explanation}, we compare the performance of explanation generation methods in two groups.
We first analyze models that make use of item features (i.e., ACMLM, NETE and PETER+).
Our PETER+ consistently and significantly outperforms ACMLM and NETE on the three datasets in terms of text quality (BLEU and ROUGE).
This shows the effectiveness of our model in generating high-quality sentences.
Notice that \citet{WWW20-NETE} conducted a user survey and reported that NETE's explanations were perceived useful by most participants.
It suggests that our model's explanations with better quality could also be very useful to real users.

Again, in terms of text quality, the performance gap between PETER+ and ACMLM (a fine-tuned BERT) is extremely large, because the latter's generation is achieved by predicting masked tokens, which is quite different from word-by-word generation.
This may explain why ACMLM produces diverse sentences (high USR), which, however, is less meaningful when text quality cannot be guaranteed.
Furthermore, PETER+ beats both ACMLM and NETE on the explainability metric FMR that cares about whether a generated explanation mentions the feature in the ground-truth.
This is quite useful in real-world applications when the system is asked to explain a particular feature.
Regarding the other two explainability metrics FCR and DIV, PETER+ is also very competitive.
ACMLM gains better performance on some cases, because at the training stage it is exposed to more features (from both the user and the item), which is unfair to both PETER+ and NETE.

Next, we discuss the results of the models that only leverage user and item IDs for generation.
As it can be seen, Transformer generates identical explanations on each dataset, resulting in nearly 0 score on Unique Sentence Ratio (USR).
Owing to the context prediction task, our PETER successfully addresses this issue, producing diverse (comparable USR) and high-quality (best BLEU-4) sentences.
In particular, on the largest dataset Yelp, it achieves the best performance on most of the metrics.
This again demonstrates the effectiveness of our model.
On Amazon and TripAdvisor, NRT and Att2Seq are very competitive, because we fixed their generation issues (see Section \ref{sec:methods}).
In addition, the two datasets are small and thus the training samples are limited, so our model may underfit, which is why it does not always reach the best performance.

\begin{table}
	\centering
	\small
	\begin{tabular}{r|cc|cc|cc}
		\hline
		\multicolumn{1}{c|}{\multirow{2}{*}{}} & \multicolumn{2}{c|}{Yelp} & \multicolumn{2}{c|}{Amazon} & \multicolumn{2}{c}{TripAdvisor} \\ \cline{2-7}
		\multicolumn{1}{c|}{}  & R$\downarrow$ & M$\downarrow$ & R$\downarrow$ & M$\downarrow$ & R$\downarrow$ & M$\downarrow$ \\ \hline
		PMF & 1.09 & 0.88 & 1.03 & 0.81 & 0.87 & 0.70 \\
		SVD++ & \textbf{1.01} & \textbf{0.78} & 0.96 & 0.72 & 0.80 & 0.61 \\
		NRT & \textbf{1.01} & \textbf{0.78} & \textbf{0.95} & \textbf{0.70} & \textbf{0.79} & 0.61 \\
		NETE & \textbf{1.01} & 0.79 & 0.96 & 0.73 & \textbf{0.79} & \textbf{0.60} \\
		PETER & \textbf{1.01} & \textbf{0.78} & \textbf{0.95} & 0.71 & 0.81 & 0.63 \\ \hline
	\end{tabular}
	\caption{Recommendation performance comparison in terms of RMSE (R for short) and MAE (denoted as M). The best performing values are boldfaced.}
	\vspace{-10pt}
	\label{table:recommendation}
\end{table}

\begin{table*}
	\centering
	\small
	\begin{tabular}{l|lll|lll|ll}
		\hline
		\multirow{2}{*}{} & \multicolumn{3}{c|}{Explainability} & \multicolumn{3}{c|}{Text Quality}  & \multicolumn{2}{c}{Recommendation} \\ \cline{2-9}
		& FMR & FCR & DIV & USR & BLEU-1 & BLEU-4 & RMSE & MAE \\ \hline		
		Disable $\mathcal{L}_c$ & 0.06 $\downarrow$ & 0.03 $\downarrow$ & 5.75 $\downarrow$ & 0.01 $\downarrow$ & 15.37 $\downarrow$ & 0.86 $\downarrow$ & 0.80 $\uparrow$ & 0.61 $\uparrow$ \\
		Disable $\mathcal{L}_r$ & 0.07 & 0.14 $\uparrow$ & 2.90 $\uparrow$ & 0.10 $\uparrow$ & 16.16 $\uparrow$ & 1.15 $\uparrow$ & 3.23 $\downarrow$ & 3.10 $\downarrow$ \\
		Left-to-Right Masking & 0.07 & 0.15 $\uparrow$ & 2.68 $\uparrow$ & 0.12 $\uparrow$ & 15.73 $\downarrow$ & 1.11 & 0.87 $\downarrow$ & 0.68 $\downarrow$ \\ \hline
		PETER & 0.07 & 0.13 & 2.95 & 0.08 & 15.96 & 1.11 & 0.81 & 0.63 \\ \hline
	\end{tabular}
	\caption{Ablation study on the smallest dataset TripAdvisor. Arrows $\uparrow$ and $\downarrow$ respectively denote the performance increase and decrease compared with PETER.}
	\vspace{-10pt}
	\label{table:ablation}
\end{table*}

Besides explanation performance, we also investigate the efficiency of different Transformer-based models.
On the same machine (NVIDIA Tesla P40) and dataset (TripAdvisor), we compare the training minutes of ACMLM and our PETER+ in Table \ref{table:efficiency}.
Compared with ACMLM, our model takes less time to train (2.3 minutes per epoch), since it has only 2 layers and thus less parameters.
But because it is unpretrained and learned from scratch, it needs more training epochs.

\subsection{Qualitative Case Study on Explanations}

In Table \ref{table:case}, we present two examples generated by PETER and PETER+ on the TripAdvisor dataset.
We can see that PETER generates distinct context words and explanations for different user-item pairs.
This confirms that our proposed solution can indeed endow the user and item IDs with linguistic meanings, as well as achieving certain degree of personalization for natural language generation.
Among the commonly used context words, e.g., \textit{the}, there are some important features (underlined), according to which the model then generates an explanation that talks about them.
Admittedly, there is still much room for improvement of the context prediction task, so as to more accurately predict the features in the ground-truth (e.g., \textit{rooms} vs. \textit{pool} in the first example).
One alternative is to leverage the features to guide the model's generation.
This explains why PETER+ is able to generate an explanation that talks about \textit{rooms} rather than \textit{pool}, making it semantically closer to the ground-truth.
It thus demonstrates our model's flexibility in incorporating these features.

\subsection{Recommendation Performance}

Table \ref{table:recommendation} presents the performance comparison of different recommendation methods.
On the largest dataset Yelp with approximately 1.3 million records, our model PETER performs as good as the three competitive baselines (i.e., SVD++, NRT and NETE), which shows the rationale of our recommendation module.
Since our model PETER has more parameters to learn, it may underfit on small datasets.
This explains why it does not always perform the best on TripAdvisor and Amazon.
When more training data are available to Transformer, usually the performance will become better, as evidenced by GPT-2 \citep{19-GPT2} and GPT-3 \citep{NeurIPS20-GPT3}.
Thus, we can expect our model to perform well in real-world applications, where the training data are bigger than the testing datasets, e.g., billion-scale users in Amazon.

\subsection{Ablation Study}

In Table \ref{table:ablation}, we provide an ablation study conducted on the TripAdvisor dataset.
After disabling the context prediction task $\mathcal{L}_c$ by setting $\lambda_c = 0$, the performances of both explainability and text quality drop dramatically, and the unique sentence ratio (USR) is nearly approaching Transformer's (see Table \ref{table:explanation}).
It hence confirms this task's effectiveness.
As $\mathcal{L}_c$ is highly correlated with the recommendation task $\mathcal{L}_r$ via the user and item IDs (see Section \ref{sec:task}), the removal of $\mathcal{L}_c$ leads to slight improvement on recommendation performance.
We can also observe a reversed phenomenon when we disable $\mathcal{L}_r$.
When PETER masking is replaced by the Left-to-Right masking that prevents the model from accessing the item information, the recommendation performance drops sharply.
Overall, PETER reaches an optimal situation, where its explainability, text quality and recommendation performance are all reasonably good.

\section{Conclusion}

We propose a simple and effective solution to address the personalized generation problem of Transformer, unleashing its language modeling power to generate explanations for recommender systems.
Extensive experiments show that the solution is both effective and efficient.
It opens up a new way of exploiting Transformer by designing good tasks instead of scaling up model size.
There are various applications of personalized generation for which Transformer is still less explored.
Our next step is to adopt our solution for personalized question answering systems and personalized conversational agents.
We also plan to incorporate item images into the model, so as to generate visual explanations for recommendations, since ``a picture is worth a thousand words''.
Another meaningful extension is to adapt the model to cross-lingual explanation generation, because international platforms, e.g., Amazon, may serve users who speak different languages.

\section*{Acknowledgments}
This work was partially supported by HKBU IRCMS/19-20/D05, RGC/HKBU12201620, and NSF IIS-1910154 and IIS-2007907.
Any opinions, findings, conclusions or recommendations expressed in this material are those of the authors and do not necessarily reflect those of the sponsors.

\bibliographystyle{acl_natbib}
\bibliography{bibliography}

\begin{thebibliography}{42}
\expandafter\ifx\csname natexlab\endcsname\relax\def\natexlab#1{#1}\fi

\bibitem[{Ai et~al.(2018)Ai, Azizi, Chen, and Zhang}]{MDPI18-KG}
Qingyao Ai, Vahid Azizi, Xu~Chen, and Yongfeng Zhang. 2018.
\newblock Learning heterogeneous knowledge base embeddings for explainable
  recommendation.
\newblock \emph{Algorithms}, 11(9):137.

\bibitem[{Brown et~al.(2020)Brown, Mann, Ryder, Subbiah, Kaplan, Dhariwal,
  Neelakantan, Shyam, Sastry, Askell et~al.}]{NeurIPS20-GPT3}
Tom~B Brown, Benjamin Mann, Nick Ryder, Melanie Subbiah, Jared Kaplan, Prafulla
  Dhariwal, Arvind Neelakantan, Pranav Shyam, Girish Sastry, Amanda Askell,
  et~al. 2020.
\newblock Language models are few-shot learners.
\newblock In \emph{Advances in neural information processing systems}.

\bibitem[{Chen et~al.(2019{\natexlab{a}})Chen, Chen, Shi, and
  Zhang}]{EARS19-HSS}
Hanxiong Chen, Xu~Chen, Shaoyun Shi, and Yongfeng Zhang. 2019{\natexlab{a}}.
\newblock Generate natural language explanations for recommendation.
\newblock In \emph{Proceedings of SIGIR'19 Workshop on ExplainAble
  Recommendation and Search}. ACM.

\bibitem[{Chen et~al.(2021)Chen, Shi, Li, and Zhang}]{chen2021neural}
Hanxiong Chen, Shaoyun Shi, Yunqi Li, and Yongfeng Zhang. 2021.
\newblock Neural collaborative reasoning.
\newblock In \emph{Proceedings of The Web Conference 2021}.

\bibitem[{Chen and Wang(2017)}]{IUI17-HCI}
Li~Chen and Feng Wang. 2017.
\newblock Explaining recommendations based on feature sentiments in product
  reviews.
\newblock In \emph{Proceedings of the 22nd International Conference on
  Intelligent User Interfaces}, pages 17--28.

\bibitem[{Chen et~al.(2019{\natexlab{b}})Chen, Yan, and Wang}]{chen2019user}
Li~Chen, Dongning Yan, and Feng Wang. 2019{\natexlab{b}}.
\newblock User evaluations on sentiment-based recommendation explanations.
\newblock \emph{ACM Transactions on Interactive Intelligent Systems (TiiS)},
  9(4):1--38.

\bibitem[{Chen et~al.(2019{\natexlab{c}})Chen, Chen, Xu, Zhang, Cao, Qin, and
  Zha}]{SIGIR19-VECF}
Xu~Chen, Hanxiong Chen, Hongteng Xu, Yongfeng Zhang, Yixin Cao, Zheng Qin, and
  Hongyuan Zha. 2019{\natexlab{c}}.
\newblock Personalized fashion recommendation with visual explanations based on
  multimodal attention network: Towards visually explainable recommendation.
\newblock In \emph{Proceedings of the 42nd International ACM SIGIR Conference
  on Research and Development in Information Retrieval}, pages 765--774.

\bibitem[{Chen et~al.(2019{\natexlab{d}})Chen, Zhang, and Qin}]{AAAI19-DER}
Xu~Chen, Yongfeng Zhang, and Zheng Qin. 2019{\natexlab{d}}.
\newblock Dynamic explainable recommendation based on neural attentive models.
\newblock In \emph{Proceedings of the AAAI Conference on Artificial
  Intelligence}, volume~33, pages 53--60.

\bibitem[{Chen et~al.(2020)Chen, Wang, Xie, Parsana, Soni, Ao, and
  Chen}]{IJCAI20-ECR}
Zhongxia Chen, Xiting Wang, Xing Xie, Mehul Parsana, Akshay Soni, Xiang Ao, and
  Enhong Chen. 2020.
\newblock Towards explainable conversational recommendation.
\newblock In \emph{Proceedings of the Twenty-Ninth International Joint
  Conference on Artificial Intelligence}.

\bibitem[{Cho et~al.(2014)Cho, Van~Merri{\"e}nboer, Gulcehre, Bahdanau,
  Bougares, Schwenk, and Bengio}]{EMNLP14-GRU}
Kyunghyun Cho, Bart Van~Merri{\"e}nboer, Caglar Gulcehre, Dzmitry Bahdanau,
  Fethi Bougares, Holger Schwenk, and Yoshua Bengio. 2014.
\newblock Learning phrase representations using rnn encoder-decoder for
  statistical machine translation.
\newblock In \emph{Proceedings of the 2014 conference on empirical methods in
  natural language processing (EMNLP)}, pages 1724--1734.

\bibitem[{Devlin et~al.(2019)Devlin, Chang, Lee, and Toutanova}]{NAACL19-BERT}
Jacob Devlin, Ming-Wei Chang, Kenton Lee, and Kristina Toutanova. 2019.
\newblock Bert: Pre-training of deep bidirectional transformers for language
  understanding.
\newblock In \emph{2019 Annual Conference of the North American Chapter of the
  Association for Computational Linguistics}.

\bibitem[{Dong et~al.(2017)Dong, Huang, Wei, Lapata, Zhou, and
  Xu}]{EACL17-Att2Seq}
Li~Dong, Shaohan Huang, Furu Wei, Mirella Lapata, Ming Zhou, and Ke~Xu. 2017.
\newblock Learning to generate product reviews from attributes.
\newblock In \emph{Proceedings of the 15th Conference of the European Chapter
  of the Association for Computational Linguistics: Volume 1, Long Papers},
  pages 623--632.

\bibitem[{Dong et~al.(2019)Dong, Yang, Wang, Wei, Liu, Wang, Gao, Zhou, and
  Hon}]{NeurIPS19-UNILM}
Li~Dong, Nan Yang, Wenhui Wang, Furu Wei, Xiaodong Liu, Yu~Wang, Jianfeng Gao,
  Ming Zhou, and Hsiao-Wuen Hon. 2019.
\newblock Unified language model pre-training for natural language
  understanding and generation.
\newblock In \emph{Advances in Neural Information Processing Systems}, pages
  13063--13075.

\bibitem[{Fu et~al.(2020)Fu, Xian, Gao, Zhao, Huang, Ge, Xu, Geng, Shah, Zhang
  et~al.}]{SIGIR20-KGAT}
Zuohui Fu, Yikun Xian, Ruoyuan Gao, Jieyu Zhao, Qiaoying Huang, Yingqiang Ge,
  Shuyuan Xu, Shijie Geng, Chirag Shah, Yongfeng Zhang, et~al. 2020.
\newblock Fairness-aware explainable recommendation over knowledge graphs.
\newblock In \emph{Proceedings of the 43rd International ACM SIGIR Conference
  on Research and Development in Information Retrieval}.

\bibitem[{Gedikli et~al.(2014)Gedikli, Jannach, and Ge}]{IJHCS14-HCI}
Fatih Gedikli, Dietmar Jannach, and Mouzhi Ge. 2014.
\newblock How should i explain? a comparison of different explanation types for
  recommender systems.
\newblock \emph{International Journal of Human-Computer Studies},
  72(4):367--382.

\bibitem[{Hochreiter and Schmidhuber(1997)}]{Neural97-LSTM}
Sepp Hochreiter and J{\"u}rgen Schmidhuber. 1997.
\newblock Long short-term memory.
\newblock \emph{Neural computation}, 9(8):1735--1780.

\bibitem[{Koren(2008)}]{KDD08-SVDpp}
Yehuda Koren. 2008.
\newblock Factorization meets the neighborhood: a multifaceted collaborative
  filtering model.
\newblock In \emph{Proceedings of the 14th ACM SIGKDD international conference
  on Knowledge discovery and data mining}, pages 426--434.

\bibitem[{Li et~al.(2019)Li, Li, and Zong}]{AAAI19-USN}
Junjie Li, Haoran Li, and Chengqing Zong. 2019.
\newblock Towards personalized review summarization via user-aware sequence
  network.
\newblock In \emph{Proceedings of the AAAI Conference on Artificial
  Intelligence}, volume~33, pages 6690--6697.

\bibitem[{Li et~al.(2020{\natexlab{a}})Li, Chen, and Dong}]{JIIS20-CAESAR}
Lei Li, Li~Chen, and Ruihai Dong. 2020{\natexlab{a}}.
\newblock Caesar: context-aware explanation based on supervised attention for
  service recommendations.
\newblock \emph{Journal of Intelligent Information Systems}, pages 1--24.

\bibitem[{Li et~al.(2020{\natexlab{b}})Li, Chen, and Zhang}]{WWW20-NETE}
Lei Li, Li~Chen, and Yongfeng Zhang. 2020{\natexlab{b}}.
\newblock Towards controllable explanation generation for recommender systems
  via neural template.
\newblock In \emph{Companion Proceedings of the Web Conference 2020}, pages
  198--202.

\bibitem[{Li et~al.(2020{\natexlab{c}})Li, Zhang, and Chen}]{CIKM20-NETE}
Lei Li, Yongfeng Zhang, and Li~Chen. 2020{\natexlab{c}}.
\newblock Generate neural template explanations for recommendation.
\newblock In \emph{Proceedings of the 29th ACM International Conference on
  Information \& Knowledge Management}, pages 755--764.

\bibitem[{Li et~al.(2021)Li, Zhang, and Chen}]{SIGIR21-EXTRA}
Lei Li, Yongfeng Zhang, and Li~Chen. 2021.
\newblock Extra: Explanation ranking datasets for explainable recommendation.
\newblock In \emph{Proceedings of the 44th International ACM SIGIR conference
  on Research and Development in Information Retrieval}.

\bibitem[{Li et~al.(2017)Li, Wang, Ren, Bing, and Lam}]{SIGIR17-NRT}
Piji Li, Zihao Wang, Zhaochun Ren, Lidong Bing, and Wai Lam. 2017.
\newblock Neural rating regression with abstractive tips generation for
  recommendation.
\newblock In \emph{Proceedings of the 40th International ACM SIGIR conference
  on Research and Development in Information Retrieval}, pages 345--354.

\bibitem[{Lin(2004)}]{TS04-ROUGE}
Chin-Yew Lin. 2004.
\newblock Rouge: A package for automatic evaluation of summaries.
\newblock In \emph{Text summarization branches out}, pages 74--81.

\bibitem[{Liu et~al.(2018)Liu, Saleh, Pot, Goodrich, Sepassi, Kaiser, and
  Shazeer}]{ICLR18-Decoder}
Peter~J Liu, Mohammad Saleh, Etienne Pot, Ben Goodrich, Ryan Sepassi, Lukasz
  Kaiser, and Noam Shazeer. 2018.
\newblock Generating wikipedia by summarizing long sequences.
\newblock In \emph{The Sixth International Conference on Learning
  Representations}.

\bibitem[{Mnih and Salakhutdinov(2007)}]{NIPS08-PMF}
Andriy Mnih and Russ~R Salakhutdinov. 2007.
\newblock Probabilistic matrix factorization.
\newblock In \emph{Advances in neural information processing systems}, pages
  1257--1264.

\bibitem[{Ni et~al.(2019)Ni, Li, and McAuley}]{EMNLP19-ACMLM}
Jianmo Ni, Jiacheng Li, and Julian McAuley. 2019.
\newblock Justifying recommendations using distantly-labeled reviews and
  fine-grained aspects.
\newblock In \emph{Proceedings of the 2019 Conference on Empirical Methods in
  Natural Language Processing and the 9th International Joint Conference on
  Natural Language Processing (EMNLP-IJCNLP)}, pages 188--197.

\bibitem[{Papineni et~al.(2002)Papineni, Roukos, Ward, and Zhu}]{ACL02-BLEU}
Kishore Papineni, Salim Roukos, Todd Ward, and Wei-Jing Zhu. 2002.
\newblock Bleu: a method for automatic evaluation of machine translation.
\newblock In \emph{Proceedings of the 40th annual meeting of the Association
  for Computational Linguistics}, pages 311--318.

\bibitem[{Pascanu et~al.(2013)Pascanu, Mikolov, and Bengio}]{ICML13-Clipping}
Razvan Pascanu, Tomas Mikolov, and Yoshua Bengio. 2013.
\newblock On the difficulty of training recurrent neural networks.
\newblock In \emph{International conference on machine learning}, pages
  1310--1318.

\bibitem[{Radford et~al.(2018)Radford, Narasimhan, Salimans, and
  Sutskever}]{18-GPT}
Alec Radford, Karthik Narasimhan, Tim Salimans, and Ilya Sutskever. 2018.
\newblock Improving language understanding by generative pre-training.

\bibitem[{Radford et~al.(2019)Radford, Wu, Child, Luan, Amodei, and
  Sutskever}]{19-GPT2}
Alec Radford, Jeffrey Wu, Rewon Child, David Luan, Dario Amodei, and Ilya
  Sutskever. 2019.
\newblock Language models are unsupervised multitask learners.
\newblock \emph{OpenAI blog}, 1(8):9.

\bibitem[{Robbins and Monro(1951)}]{TAMS51-SGD}
Herbert Robbins and Sutton Monro. 1951.
\newblock A stochastic approximation method.
\newblock \emph{The annals of mathematical statistics}, pages 400--407.

\bibitem[{Shi et~al.(2020)Shi, Chen, Ma, Mao, Zhang, and Zhang}]{shi2020neural}
Shaoyun Shi, Hanxiong Chen, Weizhi Ma, Jiaxin Mao, Min Zhang, and Yongfeng
  Zhang. 2020.
\newblock Neural logic reasoning.
\newblock In \emph{Proceedings of the 29th ACM International Conference on
  Information \& Knowledge Management}, pages 1365--1374.

\bibitem[{Vaswani et~al.(2017)Vaswani, Shazeer, Parmar, Uszkoreit, Jones,
  Gomez, Kaiser, and Polosukhin}]{NIPS17-Transformer}
Ashish Vaswani, Noam Shazeer, Niki Parmar, Jakob Uszkoreit, Llion Jones,
  Aidan~N Gomez, {\L}ukasz Kaiser, and Illia Polosukhin. 2017.
\newblock Attention is all you need.
\newblock In \emph{Advances in neural information processing systems}, pages
  5998--6008.

\bibitem[{Xian et~al.(2019)Xian, Fu, Muthukrishnan, De~Melo, and
  Zhang}]{SIGIR19-PGPR}
Yikun Xian, Zuohui Fu, S~Muthukrishnan, Gerard De~Melo, and Yongfeng Zhang.
  2019.
\newblock Reinforcement knowledge graph reasoning for explainable
  recommendation.
\newblock In \emph{Proceedings of the 42nd International ACM SIGIR Conference
  on Research and Development in Information Retrieval}, pages 285--294.

\bibitem[{Xian et~al.(2020)Xian, Fu, Zhao, Ge, Chen, Huang, Geng, Qin, De~Melo,
  Muthukrishnan et~al.}]{CIKM20-CAFE}
Yikun Xian, Zuohui Fu, Handong Zhao, Yingqiang Ge, Xu~Chen, Qiaoying Huang,
  Shijie Geng, Zhou Qin, Gerard De~Melo, Shan Muthukrishnan, et~al. 2020.
\newblock Cafe: Coarse-to-fine neural symbolic reasoning for explainable
  recommendation.
\newblock In \emph{Proceedings of the 29th ACM International Conference on
  Information \& Knowledge Management}, pages 1645--1654.

\bibitem[{Zhang and Chen(2020)}]{FTIR20-Survey}
Yongfeng Zhang and Xu~Chen. 2020.
\newblock Explainable recommendation: A survey and new perspectives.
\newblock \emph{Foundations and Trends{\textregistered} in Information
  Retrieval}, 14(1):1--101.

\bibitem[{Zhang et~al.(2018)Zhang, Chen, Ai, Yang, and Croft}]{CIKM18-SAUR}
Yongfeng Zhang, Xu~Chen, Qingyao Ai, Liu Yang, and W~Bruce Croft. 2018.
\newblock Towards conversational search and recommendation: System ask, user
  respond.
\newblock In \emph{Proceedings of the 27th ACM International Conference on
  Information and Knowledge Management}, pages 177--186.

\bibitem[{Zhang et~al.(2014)Zhang, Lai, Zhang, Zhang, Liu, and
  Ma}]{SIGIR14-EFM}
Yongfeng Zhang, Guokun Lai, Min Zhang, Yi~Zhang, Yiqun Liu, and Shaoping Ma.
  2014.
\newblock Explicit factor models for explainable recommendation based on
  phrase-level sentiment analysis.
\newblock In \emph{Proceedings of the 37th international ACM SIGIR conference
  on Research \& development in information retrieval}, pages 83--92.

\bibitem[{Zheng et~al.(2020)Zheng, Zhang, Huang, and Mao}]{AAAI20-Chatbot}
Yinhe Zheng, Rongsheng Zhang, Minlie Huang, and Xiaoxi Mao. 2020.
\newblock A pre-training based personalized dialogue generation model with
  persona-sparse data.
\newblock In \emph{Proceedings of the AAAI Conference on Artificial
  Intelligence}, pages 9693--9700.

\bibitem[{Zhou et~al.(2020)Zhou, Zhao, Bian, Zhou, Wen, and Yu}]{KDD20-Chatbot}
Kun Zhou, Wayne~Xin Zhao, Shuqing Bian, Yuanhang Zhou, Ji-Rong Wen, and
  Jingsong Yu. 2020.
\newblock Improving conversational recommender systems via knowledge graph
  based semantic fusion.
\newblock In \emph{Proceedings of the 26th ACM SIGKDD International Conference
  on Knowledge Discovery \& Data Mining}, pages 1006--1014.

\bibitem[{Zhu et~al.(2021)Zhu, Xian, Fu, de~Melo, and
  Zhang}]{zhu2021faithfully}
Yaxin Zhu, Yikun Xian, Zuohui Fu, Gerard de~Melo, and Yongfeng Zhang. 2021.
\newblock Faithfully explainable recommendation via neural logic reasoning.
\newblock In \emph{2021 Annual Conference of the North American Chapter of the
  Association for Computational Linguistics}.

\end{thebibliography}


\end{document}